**PAPER • OPEN ACCESS**

# Identifying extra high frequency gravitational waves generated from oscillons with cuspy potentials using deep neural networks

To cite this article: Li-Li Wang *et al* 2019 *New J. Phys.* **21** 043005

View the article online for updates and enhancements.

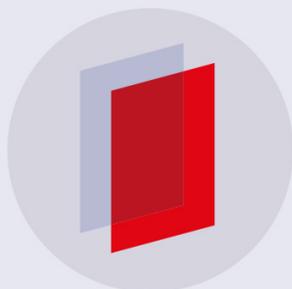

IOP Publishing    New J. Phys. 21 (2019) 043005    https://doi.org/10.1088/1367-2630/ab1310# New Journal of Physics

The open access journal at the forefront of physics

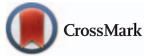

Deutsche Physikalische Gesellschaft ◆ DPG

IOP Institute of Physics

Published in partnership with: Deutsche Physikalische Gesellschaft and the Institute of Physics

PAPER

# Identifying extra high frequency gravitational waves generated from oscillons with cuspy potentials using deep neural networks

Li-Li Wang[1], Jin Li[1,4], Nan Yang[2] and Xin Li[3]

[1] Department of Physics, Chongqing University, Chongqing 401331, People's Republic of China
[2] Department of Electronical Information Science and Technology, Xingtai University, Xingtai 054001, People's Republic of China
[3] Department of Physics, Chongqing University, Chongqing 401331, People's Republic of China
[4] Author to whom any correspondence should be addressed.

E-mail: 20152702016@cqu.edu.cn, cqujinli1983@cqu.edu.cn, cqunanyang@hotmail.com and lixin1981@cqu.edu.cnCrossMark

OPEN ACCESS

RECEIVED
2 November 2018

REVISED
3 March 2019

ACCEPTED FOR PUBLICATION
25 March 2019

PUBLISHED
8 April 2019

Keywords: extra high frequency gravitational waves, deep neural networks, signal classification, parameters estimationOriginal content from this work may be used under the terms of the Creative Commons Attribution 3.0 licence.

Any further distribution of this work must maintain attribution to the author(s) and the title of the work, journal citation and DOI.

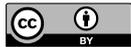## Abstract

During oscillations of cosmology inflation around the minimum of a cuspy potential after inflation, the existence of extra high frequency gravitational waves (HFGWs) (∼GHz) has been proven effectively recently. Based on the electromagnetic resonance system for detecting such extra HFGWs, we adopt a new data processing scheme to identify the corresponding GW signal, which is the transverse perturbative photon fluxes (PPF). In order to overcome the problems of low efficiency and high interference in traditional data processing methods, we adopt deep learning to extract PPF and make some source parameters estimation. Deep learning is able to provide an effective method to realize classification and prediction tasks. Meanwhile, we also adopt anti-overfitting technique and make adjustment of some hyperparameters in the course of study, which improve the performance of classifier and predictor to a certain extent. Here the convolutional neural network (CNN) is used to implement deep learning process concretely. In this case, we investigate the classification accuracy varying with the ratio between the number of positive and negative samples. When such ratio exceeds to 0.11, the accuracy could reach up to 100%. Besides, we also investigate the classification accuracy with different amplitude of extra HFGWs. As a predictor, the mean relative error of parameters estimation decreases when the amplitude of extra HFGWs increases. Especially, when amplitude $h(t)$ is in $10^{-31}$–$10^{-30}$ the mean relative error reaches around 0.014. On the contrary, the mean relative error increases with frequency increasing in $10^{8}$–$10^{11}$ Hz. At the optimal resonance frequency $5 \times 10^{9}$ Hz, the mean relative error is 0.12. Then we also study the mean relative error varying with waist radius $W_0$ of Gaussian beam, its optimal value is 0.138 when $W_0$ is in (0.05 m, 0.1 m) approximately. Compared with classifiers and predictors using other machine learning algorithms, deep CNN for our datasets has higher accuracy and lower error.## 1. Introduction

GW as one of the predictions of general relativity, has been discussed intensively in astronomy and theoretical physics. Currently several GW signals emitted from coalescence of binary black holes and binary neutron stars were verified as reality, which are all contributed to aLIGO's frequency band ($10^{1}$–$10^{3}$ Hz) [1–7]. Except those sources, GW could also arise from many other sources, including corecollapse supernovae [8], rotating neutron stars [9], coalescing stellar binaries [10–14], coalescing massive black hole binaries [15–19] and magnetars [20, 21], which are in other frequency regions. Therefore GW detectors in different frequency bands are designed and will be in operation successionally. For instance, there are pulsar timing arrays ($10^{-9}$–$10^{-7}$ Hz) [22–26], space-based interferometers such as eLISA ($10^{-4}$–$10^{0}$ Hz) [27]. In recent years, it has been indicated that inflaton oscillations around the minimum of a cuspy potential after inflation [28] and parametric resonance of $\phi$ field with other matter fields in preheating or at the end of inflation [29] could produce extra HFGWs at $10^{8}$–$10^{11}$ Hz and with dimensionless amplitude of GW $h \sim 10^{-36}$–$10^{-30}$. The source of extra HFGWs (i.e. inflaton oscillations around

© 2019 The Author(s). Published by IOP Publishing Ltd on behalf of the Institute of Physics and Deutsche Physikalische Gesellschaft

the minimum of a cuspy potential after inflation) is our study object. An EM resonance system for detecting extra HFGWs regarded as a supplement of current GW projects had been proposed by Professor Li [30]. The basic principle is the electrodynamics equations in curved spacetime [31], in which background static magnetic field in fluctuation curved spacetime could generate perturbative EM field. Thus such EM field contains GW's information and is able to react with background EM field set artificially, then generating PPF in the perpendicular direction of GW propagation.

Because the PPF can reflect the existence of extra HFGWs, it is considered as our signal. Unfortunately due to the weak amplitude of extra HFGWs, PPF is always submerged by background EM field and noise. In traditional signal processing scheme [32], a special material named fractal membranes is theoretically used to diminish transverse background photon fluxes (BPFs) in specific area to ensure enough signal to noise ratio (SNR) in this area. But it leads to a great deal of potential new electromagnetic noise. So it is difficult to extract PPF through traditional signal processing method. Here we utilize deep learning viewed as one of advanced technology in machine learning to extract the signal and estimate corresponding parameters of GW source without fractal membranes. Deep learning is more expressive than traditional methods in data analysis. It has been used widely in engineering applications and gained great achievements in recent years, such as deep generative models, machine translations, attention in deep models, one-shot learning, style transfers, deep unsupervised learning in the past few years [33–38]. In our case, each dataset is a $1 \times 101$-dimensional series describing distribution of photon energy in transverse space, where the energy on each space point is a feature. Although CNN is popular on dealing with image-related tasks, it is found that much higher sensitivities at low SNR by directly inputting raw space-series data into CNNs [39, 40]. The effect of GW is always at low SNR. Directly handling raw one-dimensional series can avoid losing information rather than converting to 2D image. Therefore in this paper, we focus on the application of CNN which is able to reduce computational cost through sharing weights and small kernels, to recognize the PPF generated by extra HFGWs submerged in the stationary gaussian white noise, shot noise, noise from the inhomogeneity of background static magnetic field.

The deep learning process could be divided into two parts : 1. In classification, the data including signal (PPF) would be extracted and sorted to be GW event, otherwise the data is classified to be noise; 2. In prediction, some parameters of GW events classified successfully would be estimated. Recently using CNN to recognize GW in aLIGO frequency band and estimate the source parameters has obtained a great success [39]. It was shown that the deep filtering outperforms conventional machine learning algorithms significantly, and the results are consistent to the ones yielded by matched filtering [39, 40]. At present, the samples are not from true events because the real world GW signal in extra high frequency band (~GHz) discussed in this paper has not be detected yet. Here we generate simulated data to do our work. The research is helpful to introduce new technology for laying the theoretical foundation to future experiment.

This paper is organized as follows. In section 2, we will introduce our simulated data in theory and the CNN we designed. Then noises concerned here are stationary gaussian white noise, shot noise and noise produced by the inhomogeneity of background magnetic field. In section 3, the effect of positive to negative sample ratio in training data sets on deep CNN's accuracy is also investigated. Through tuning the hyperparameters, the classification accuracy varying with GW amplitude is obtained. Moreover, we also compare the accuracy of deep CNN with that of other machine learning algorithms. In section 4, the ability of our CNN as a predictor is discussed through adjusting the hyperparameters. Besides, we make a comparison between deep CNN and other predictors using baseline machine learning methods for estimating parameters. Conclusions and remarks are presented in section. 5.

## 2. The EM resonant response to the extra HFGWs and obtaining datasets

The EM response system consists of background EM field (i.e. Gaussian beam (GB)) and static magnetic field [32]. The EM response process includes two stages : (1) the extra HFGWs propagating along the *z*-axis interacted with static magnetic field could generate transverse (i.e. *x*-axis and *y*-axis) first-order perturbative EM field [30, 41–44]. Certainly in fact the propagating direction of extra HFGW is not always along the *z*-axis of our observation direction, so we will add an intersection angle term in calculation, which will be discussed in equation (5); (2) when the frequency of GW $\nu_g$ equals to that of the background EM field $\nu_e$, the interaction between the transverse first-order perturbative EM field and the GB could generate transverse first-order PPF. The PPF is a physical effect of extra HFGWs, which can be observed by the photon counter on the transverse direction of extra HFGW.

In general, GB of fundamental frequency mode could be expressed as [30]

$$\psi = \frac{\psi_0}{\sqrt{1 + \left(\frac{z}{z_0}\right)^2}} \exp\left(-\frac{r^2}{W^2}\right)$$
$$\times \exp\left\{i(\kappa_e z - \omega_e t) - \tan^{-1}\frac{z}{z_0} + \frac{\kappa_e r^2}{2R} + \delta\right\}, \quad (1)$$

where $\psi_0$ denotes the amplitude of GB in electric field component. $r^2 = x^2 + y^2$, $\kappa_e = 2\pi/\lambda_e$, $\omega_e = \kappa_e c$. $\lambda_e$, $\kappa_e$, $\omega_e$ is the wavelength, wave vector and angle frequency of EM field of GB, respectively. $z_0 = \frac{\pi W_0^2}{\lambda_e}$ is the Rayleigh size of the GB. $W_0$ denotes the waist radius of GB, i.e. the minimum spot radius along the $z$-axis. It can be adjusted in region (0.05 m, 0.1 m) as appropriate. $W = W_0\sqrt{1 + \left(\frac{z}{z_0}\right)^2}$, $R = z + \frac{z_0^2}{z}$ represents the curvature radius of the wave fronts of the GB at $z$-axis. Note that $z$-axis is the symmetrical axis of the GB (i.e. the propagation axis of GB). $\delta$ denotes the phase difference between GB and the resonant components of extra HFGWs. The static magnetic filed $\bar{B}_y^0$ points along $y$-axis and is located in a fixed region. The fixed region is $-l_1 \leqslant z \leqslant l_2$ along the $z$-axis, where $l_1 = 5.7$ m and $l_2 = 0.3$ m.

Without loss of generality, GB could be treated as electric component of background EM field. Thus, we set components of background electric field $E_x^0 = \psi = \psi_x$ and $E_z^0 = 0$. Then, other components of the background EM field through Maxwell's equations could be written as

$$E_y^0 = 2x\left(\frac{1}{W^2} - \frac{i\kappa_e}{2R}\right)\int E_x^0 dy, \quad (2)$$

$$B_z^0 = \frac{i}{\omega_e}\left(\frac{\partial E_x^0}{\partial y} - \frac{\partial E_y^0}{\partial x}\right). \quad (3)$$

From [30], one can find that the PPF along the $y$-axis has no observable effect because it has the same distribution as background EM field. Therefore, in this paper we only concentrate on the PPF along $x$-axis as the transverse PPF due to its distinctly different space distribution from background EM field. Then, the BPF density pointing along $x$-axis could be represented as

$$n_x^0 = \frac{1}{\hbar\omega_e}\left\langle \frac{1}{\mu_0}(E_y^0 B_z^0)\right\rangle, \quad (4)$$

where $\mu_0$ is the vacuum permeability, and the angular bracket denotes the average over time. The expressions of $E_y^0$ and $B_z^0$ are given in (2), (3) as above.

According to the Maxwell's equations in curved spacetime and spacetime fluctuation caused by GWs, when extra HFGWs pointing along a direction, which has an interaction angle $\theta$ to $z$-axis, are immersed in static magnetic field, the transverse first-order perturbative EM field could be generated, which of electric component in extra high frequency condition could be given as follow [30]

$$E_y^{(1)} = -\frac{1}{2}h(t)\bar{B}_y^0 \kappa_g c(z + l_1)\exp[i(\kappa_g z - \omega_g t)]\cos\theta$$
$$+ \frac{i}{4}h(t)\bar{B}_y^0 c \exp[i(\kappa_g z + \omega_g t)]\cos\theta, \quad (5)$$

where $h(t)$ is amplitude of extra HFGWs. $\kappa_g$, $\omega_g$ is the wave vector and angle frequency of extra HFGWs respectively, which are equal to the wave vector $\kappa_e$ and angle frequency $\omega_e$ of background EM field. Considering the average value in whole range of $\theta$, the $E_y^{(1)}$ will have 1/2 to 1/3 coefficient by integrating against the $\theta$ term [43]. In [28], the amplitude of extra HFGWs could be distributed in region $10^{-36}$–$10^{-30}$. Background EM field coupled with the first-order perturbative EM field will generate first-order perturbative energy fluxes. Then, the first-order PPF density along the $x$-axis should be expressed as

$$n_x^1 = \frac{1}{\hbar\omega_e}\left\langle \frac{1}{\mu_0}(E_y^1 B_z^0)\right\rangle, \quad (6)$$

where the expressions of $B_z^0$ and $E_y^1$ are provided in (3), (5). Then, according to formulae (4), (6), the BPF and PPF along the $x$-axis can be obtained as follow

$$N_x^0 = \iint_{\Delta s} n_x^0 dy dz, \quad (7)$$

$$N_x^1 = \iint_{\Delta s} n_x^1 dy dz, \quad (8)$$

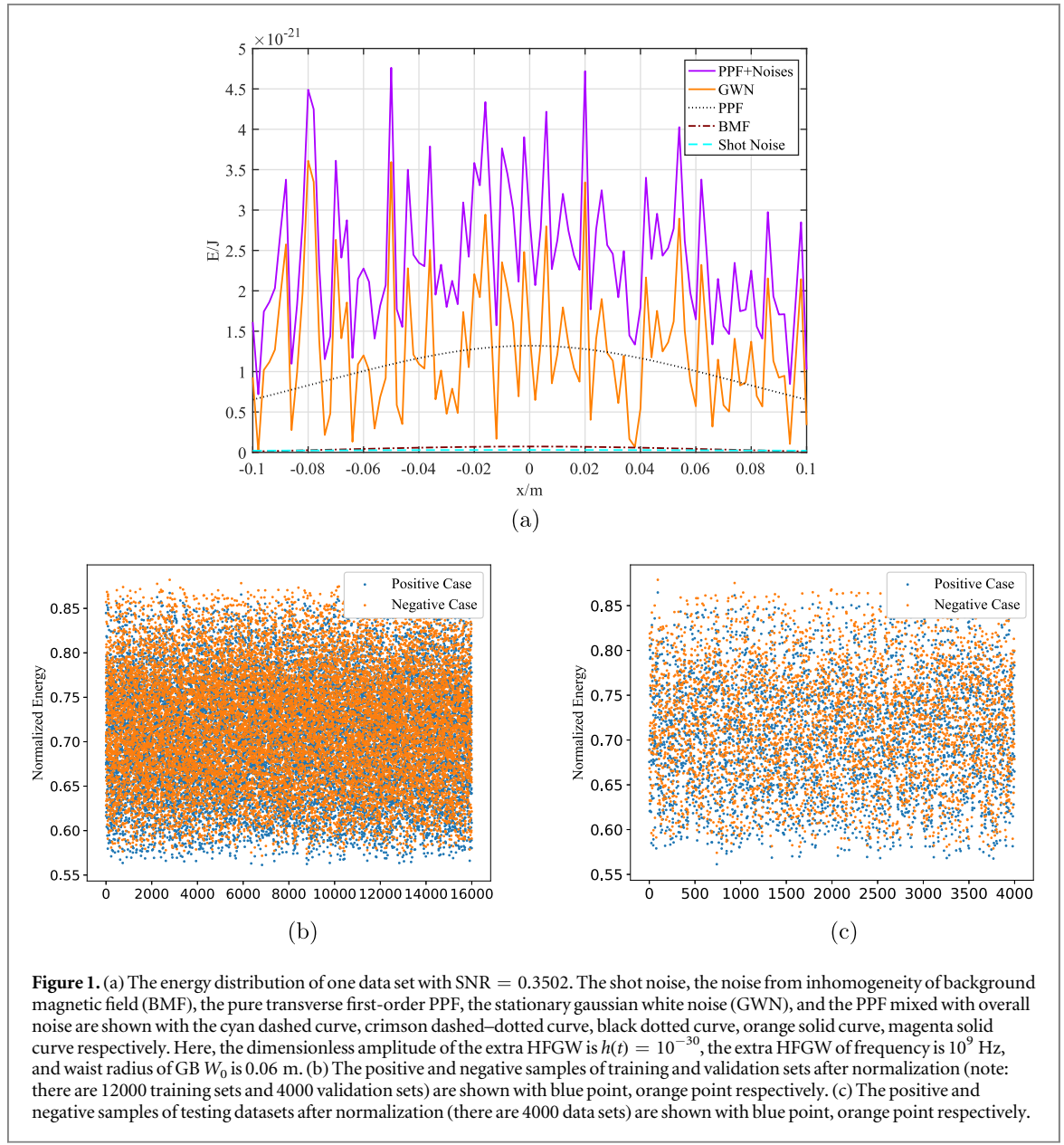

**Figure 1.** (a) The energy distribution of one data set with SNR = 0.3502. The shot noise, the noise from inhomogeneity of background magnetic field (BMF), the pure transverse first-order PPF, the stationary gaussian white noise (GWN), and the PPF mixed with overall noise are shown with the cyan dashed curve, crimson dashed–dotted curve, black dotted curve, orange solid curve, magenta solid curve respectively. Here, the dimensionless amplitude of the extra HFGW is $h(t) = 10^{-30}$, the extra HFGW of frequency is $10^9$ Hz, and waist radius of GB $W_0$ is 0.06 m. (b) The positive and negative samples of training and validation sets after normalization (note: there are 12000 training sets and 4000 validation sets) are shown with blue point, orange point respectively. (c) The positive and negative samples of testing datasets after normalization (there are 4000 data sets) are shown with blue point, orange point respectively.

where $0 < y < 0.1$ m, $0 < z < 0.3$ m denotes integral interval along *y*-axis and *z*-axis respectively. $\Delta s$ is a 'typical receiving surface' on the *yoz* plane, where the integral area is around 0.03 m$^2$.

As shown in [30], although the strength of BPF in most areas is much larger than that of PPF, they have distinct distributions. Thus the BPF could be dropped out from data by calculating difference of photons number by switching magnetic field on and off. After eliminating BPF from the interaction between background electric field and magnetic field, there could be thermal noise, shot noise [42], quantum fluctuation noise, stationary gaussian white noise and noise produced by inhomogeneous background magnetic field [45] in curve spacetime. In this paper, we take stationary gaussian white noise, shot noise and noise from inhomogeneity of background magnetic field into account. We choose parameters of the EM resonance system as follows : the power of GB $P = 10$ W, the amplitude of GB $\psi_0 \approx 1.26 \times 10^3$ V m$^{-1}$, the background static magnetic field $\bar{B}_y^0 = 10$ T, and $\delta = \frac{\pi}{2}$.

According to above principle, we can simulate data sets, which include 1. PPF with three types of noise mentioned above as positive samples, 2. pure noise as negative samples. The energy of PPF, stationary gaussian white noise, shot noise, noise from the inhomogeneity of background magnetic field and PPF mixed with overall noise are shown as figure 1(a). Compared to stationary gaussian white noise, the influence of shot noise and noise from inhomogeneous background magnetic field are not dominant. The training and testing datasets are normalized by taking the natural logarithm and dividing them by respective maximum. The positive and negative samples of training and testing sets after normalization are illuminated in figures 1(b) and (c) respectively, which shows that our positive and negative points are well mixed and would not be discriminated in

```
        Input              matrix (size: 1×101)
  1     ReshapeLayer       3–tensor (size: 1×1×101)
  2     ConvolutionLayer   3–tensor (size: 8×1×100)
  3     Ramp               3–tensor (size: 8×1×100)
  4     PoolingLayer       3–tensor (size: 8×1×100)
  5     ConvolutionLayer   3–tensor (size: 16×1×99)
  6     Ramp               3–tensor (size: 16×1×99)
  7     PoolingLayer       3–tensor (size: 16×1×99)
  8     ConvolutionLayer   3–tensor (size: 32×1×98)
  9     Ramp               3–tensor (size: 32×1×98)
 10     PoolingLayer       3–tensor (size: 32×1×98)
 11     FlattenLayer       vector (size: 3136)
 12     LinearLayer        vector (size: 30)
 13     Ramp               vector (size: 30)
 14     LinearLayer        vector (size: 2)
 15     SoftmaxLayer       vector (size: 2)
        Output             class
```

**Figure 2.** Architecture of our convolutional neural network used to do classification. The input is one of our training data sets and the output is two classes, i.e. True or False. For prediction we simply take out the softmax layer after the 15th layer and use the mean relative error as the loss function.

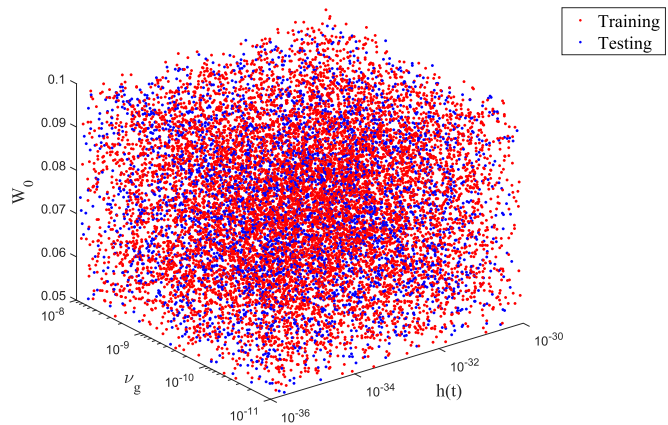

**Figure 3.** Our simulated training and testing datasets varying with the amplitude $h(t)$, frequency $\nu_g = \omega_g/2\pi$ of extra HFGWs, and the waist radius $W_0$ of GB. On Mathematica platform, the validation sets can be automatically chosen from training datasets. Here we set the ratios of training, validation and testing datasets to be 60%, 20%, 20% respectively.

an obvious way. Our aim is to recognize the data sets including PPF from the ones only containing pure noise. Firstly, we designed our deep CNN as figure 2. Secondly, in order to find the optimal weights and bias of our CNN, we put a large number of data sets called as training sets to train the CNN. Meanwhile, in the process of training data sets, we apply validation sets for anti-overfitting. Finally, the accuracy of classifier and mean relative error of predictor can be obtained using the trained deep CNN to make a judgment on testing data sets. Note: the training sets, validation sets and testing sets were chosen to be disjoint due to different source parameters as shown in figure 3. In the training process, we tune the hyperparameters according to the loss curve of training datasets and validation datasets on Mathematica platform, and choose their optimal values which result in the minimum gap between the loss of training datasets and validation datasets.

## 3. The application of deep learning for classification

Here we investigate classification accuracy of deep CNN for 12000 training data sets in amplitude region $10^{-36}$–$10^{-30}$. In order to achieve optimal results of our CNN, we spend great effort on tuning hyperparameters. In this paper, we take the accuracy as a metric measuring performance of classifier, which indicates the fraction of samples classified correctly. Using the CNN architecture as shown in figure 2, the results with various hyperparameters are shown in figure 5, where one can find that the classification accuracy of CNN with some specific hyperparameters is hopeful to be 100% in amplitude region $10^{-36}$–$10^{-30}$, i.e. noisy signals with amplitude $10^{-36}$–$10^{-30}$ could be recognized. Some anti-overfitting methods, such as DropoutLayer, L2Regularization, and other two hyperparameters: MaxTrainingRounds (i.e. number of iterations) and BatchSize (i.e. number of samples trained in a batch) are concerned. The feature maps for our deep CNN are figured out in figure 4, which shows the CNN indeed extracts the key feature of pure signal and makes effective discrimination of positive and negative samples.

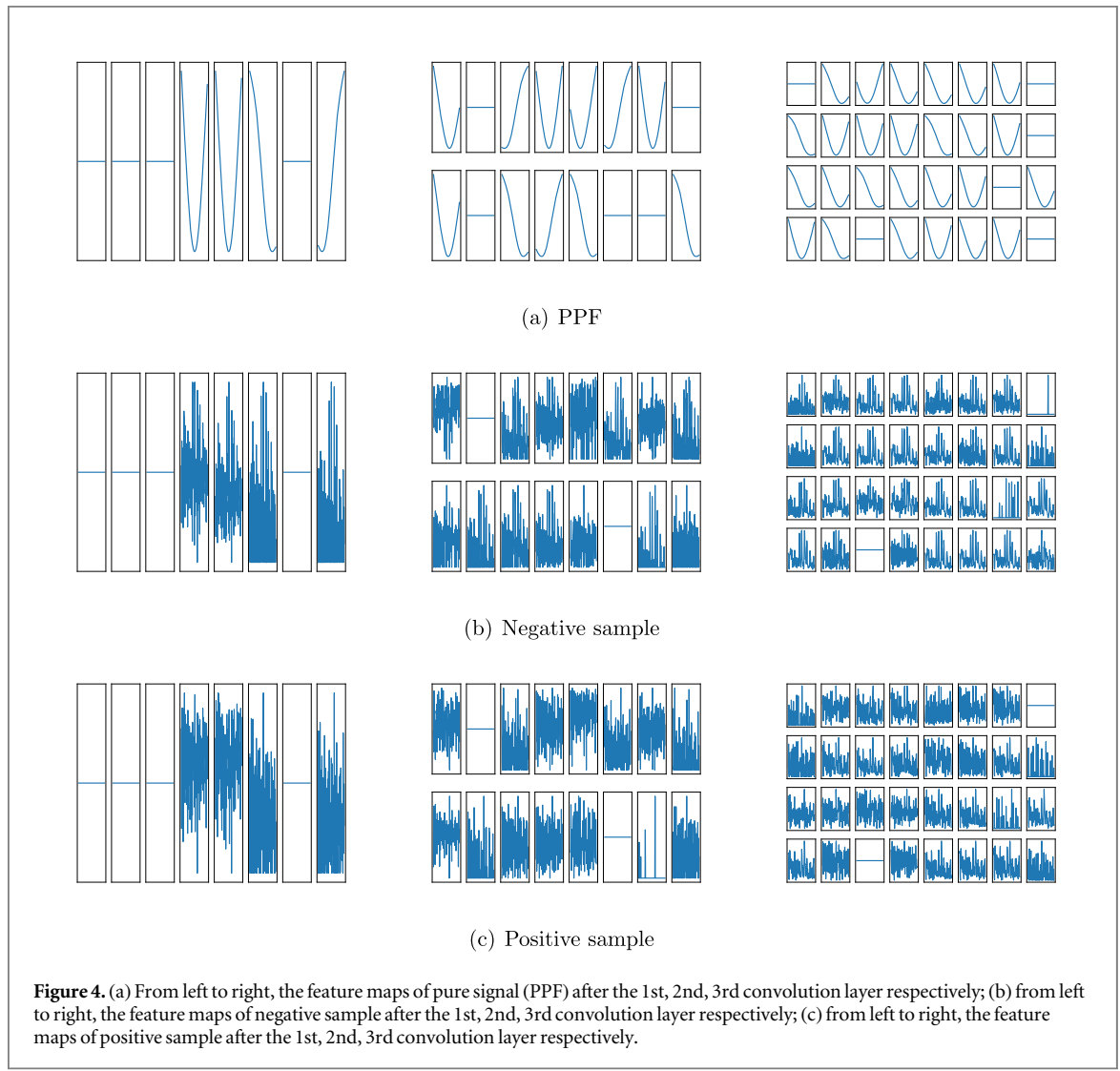

**Figure 4.** (a) From left to right, the feature maps of pure signal (PPF) after the 1st, 2nd, 3rd convolution layer respectively; (b) from left to right, the feature maps of negative sample after the 1st, 2nd, 3rd convolution layer respectively; (c) from left to right, the feature maps of positive sample after the 1st, 2nd, 3rd convolution layer respectively.

However, it is inevitable that the ratio of positive and negative samples is severely imbalanced in real world, especially GW events in high frequency band have not been detected yet. As shown in figure 6, when the ratio of position and negative samples is 0.03, the accuracy could reach up to 97.63%, and the corresponding AUC (i.e. the area under receiver operating characteristic curve) is 1. Once the ratio exceeds to 0.11, the accuracy could go up to 100%. In some research work of GW, the same number of noisy GW signals and pure noise was adopted [39, 40].

DropoutLayer and L2Regularization are the main anti-overfitting methods adopted in this paper, which could improve accuracy of deep CNN to a certain extent. In our results, from (a) in figure 5, when the DropoutLayer is set to be 0.5, the accuracy of deep CNN could reach up to 1 in amplitude range $h(t) \sim 10^{-36}$–$10^{-30}$. Thus in the following context the dropout ratio is determined to be 0.5. For L2Rrgularization, the regularization coefficient chosen to be 0 may be preferable for our case in each interval of the amplitude as shown in (b) of figure 5. Furthermore, we also study other two hyperparameters: MaxTrainingRounds and Batchsize. As shown in (c) and (d) of figure 5, when their value is set to be 150, 100 respectively, the accuracy of classifiers could reach to 1 in whole amplitude range.

As a comparison, we also investigate the performance for our data sets through commonly used machine learning methods, including Random Forest, Support Vector Machine, k-Nearest Neighbors, Neural Networks,

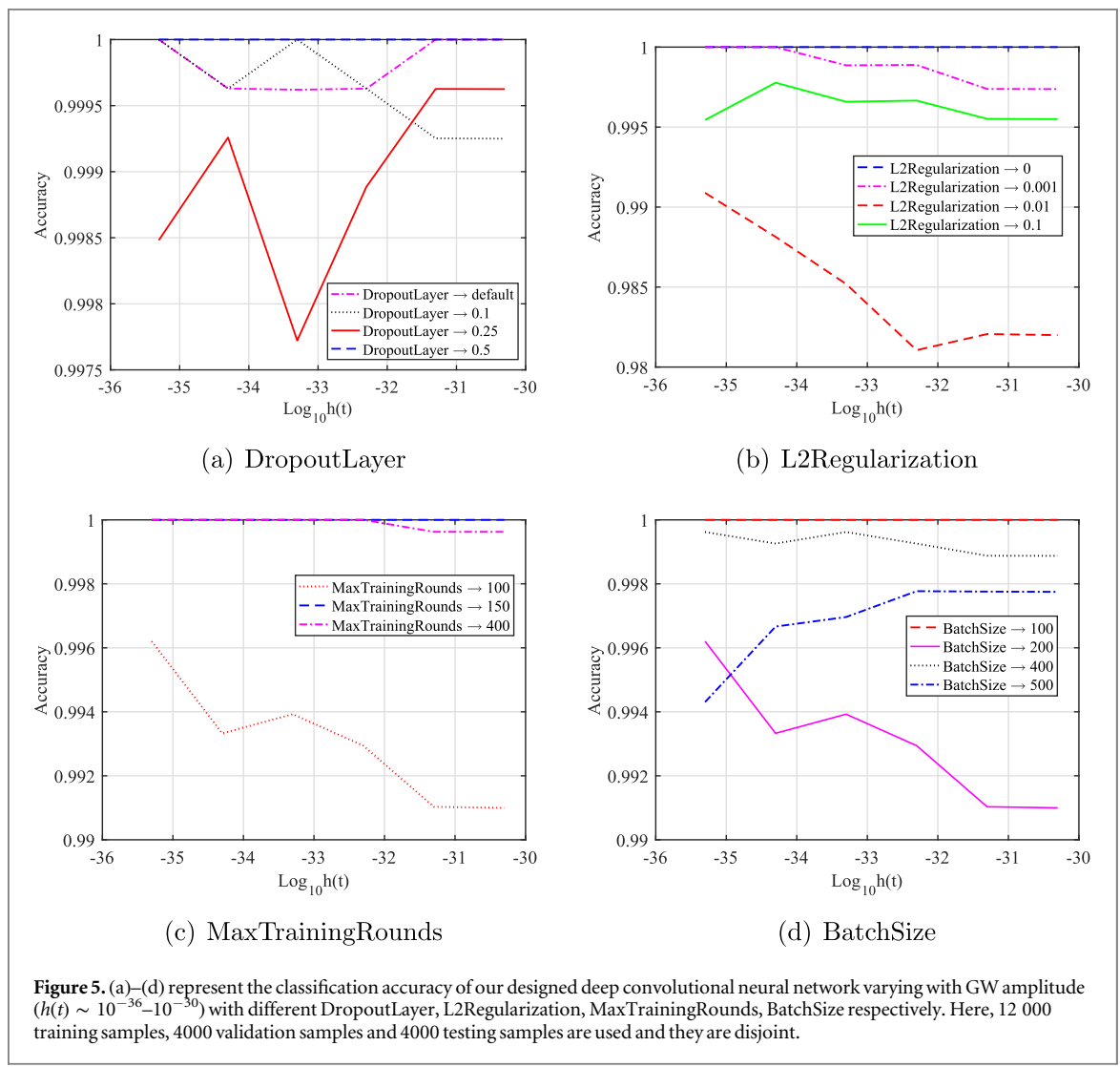

**Figure 5.** (a)–(d) represent the classification accuracy of our designed deep convolutional neural network varying with GW amplitude ($h(t) \sim 10^{-36}$–$10^{-30}$) with different DropoutLayer, L2Regularization, MaxTrainingRounds, BatchSize respectively. Here, 12 000 training samples, 4000 validation samples and 4000 testing samples are used and they are disjoint.

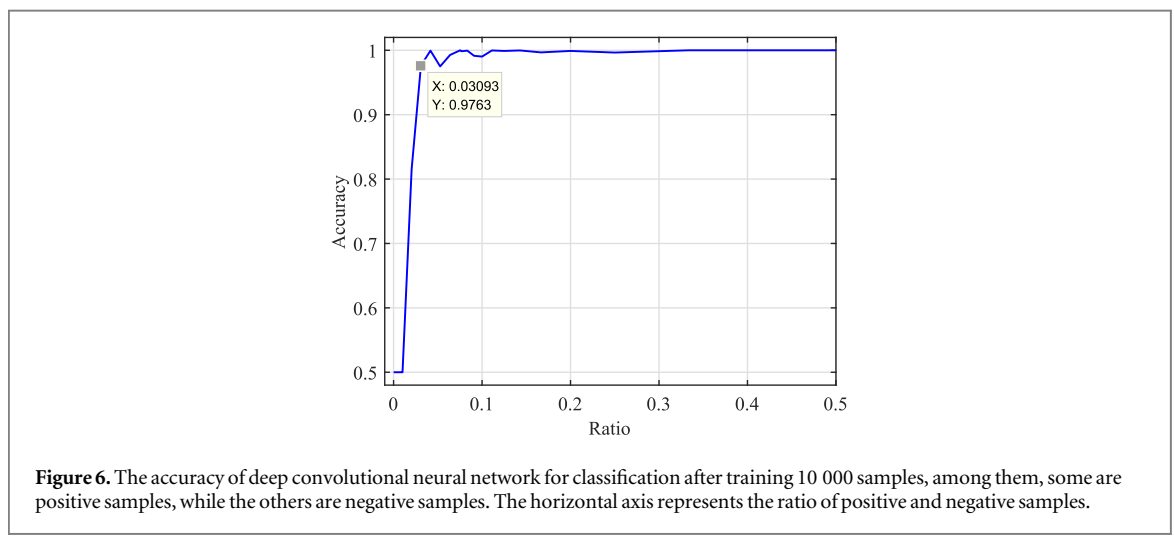

**Figure 6.** The accuracy of deep convolutional neural network for classification after training 10 000 samples, among them, some are positive samples, while the others are negative samples. The horizontal axis represents the ratio of positive and negative samples.

Logistic Regression and Naive Bayes. Here the Neural Network is a simple feedforward neural network from the input to output layer connected with one hidden layer. And for different machine learning methods, the AUC are 1, 0.5648, 0.5545, 0.5530, 0.5493, 0.5417, 0.5156 respectively for deep CNN, Naive Bayes, Logistic Regression, Neural Network, Nearest Neighbors, Support Vector Machine, Random Forest (see figure 7(b)). Therefore in our case the traditional machine learning classifiers are not so reliable as deep CNN.

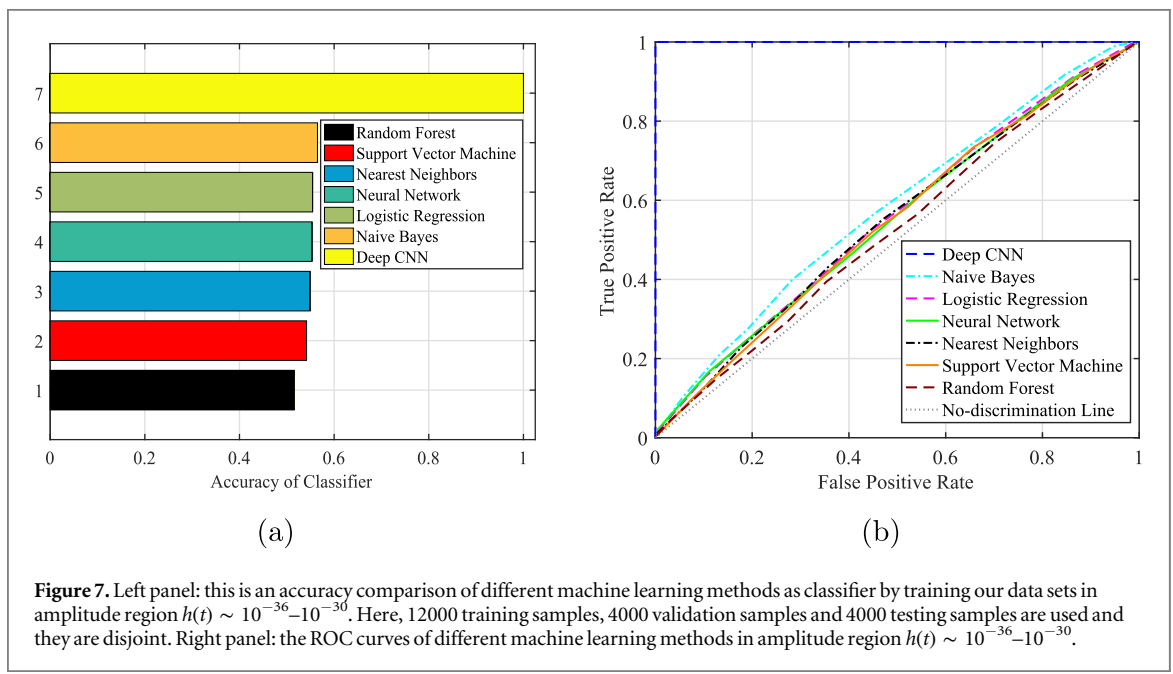

**Figure 7.** Left panel: this is an accuracy comparison of different machine learning methods as classifier by training our data sets in amplitude region $h(t) \sim 10^{-36}$–$10^{-30}$. Here, 12000 training samples, 4000 validation samples and 4000 testing samples are used and they are disjoint. Right panel: the ROC curves of different machine learning methods in amplitude region $h(t) \sim 10^{-36}$–$10^{-30}$.

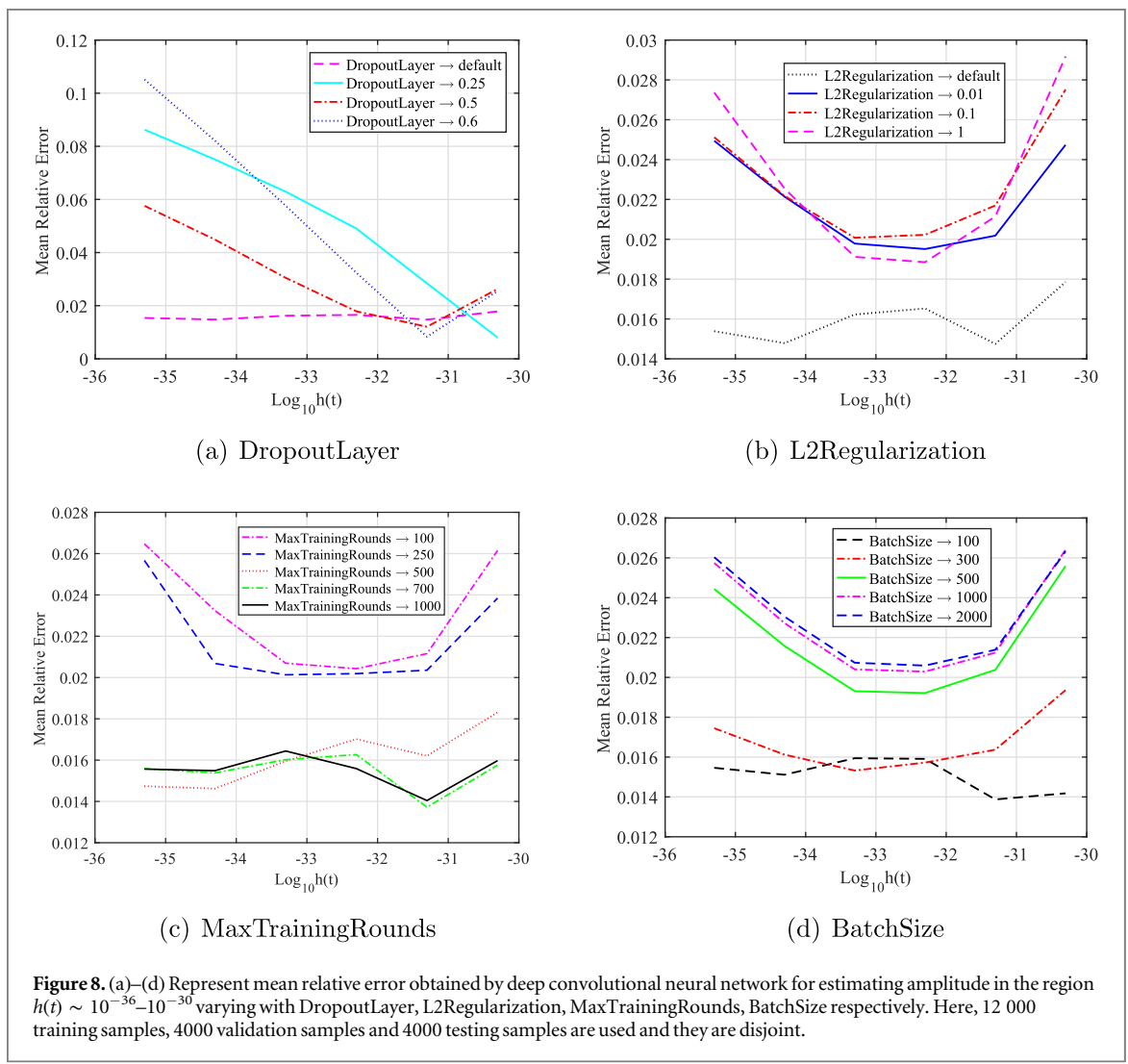

**Figure 8.** (a)–(d) Represent mean relative error obtained by deep convolutional neural network for estimating amplitude in the region $h(t) \sim 10^{-36}$–$10^{-30}$ varying with DropoutLayer, L2Regularization, MaxTrainingRounds, BatchSize respectively. Here, 12 000 training samples, 4000 validation samples and 4000 testing samples are used and they are disjoint.

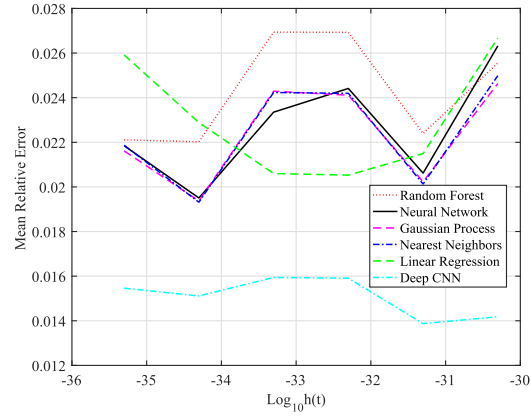

**Figure 9.** This is the mean relative error obtained by various machine learning algorithms for estimating amplitude in region $h(t) \sim 10^{-36}$–$10^{-30}$. Here, 12 000 training samples, 4000 validation samples and 4000 testing samples are used and they are disjoint.

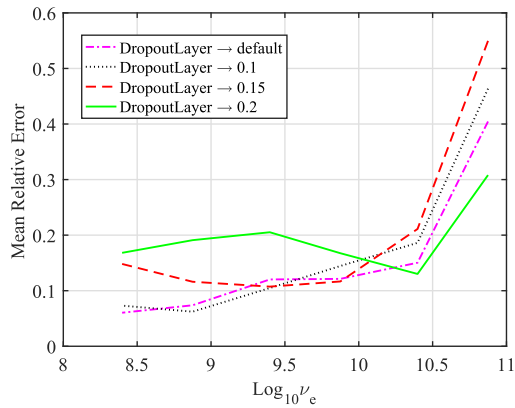

(a) DropoutLayer

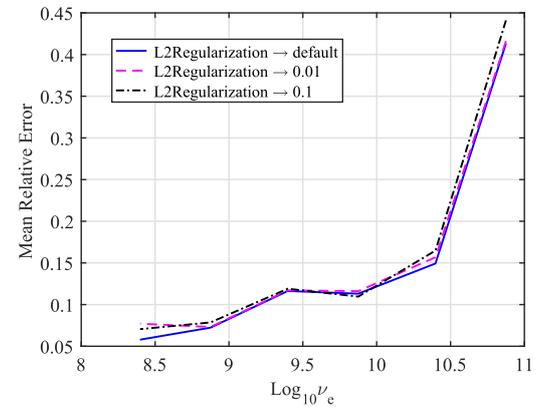

(b) L2Regularization

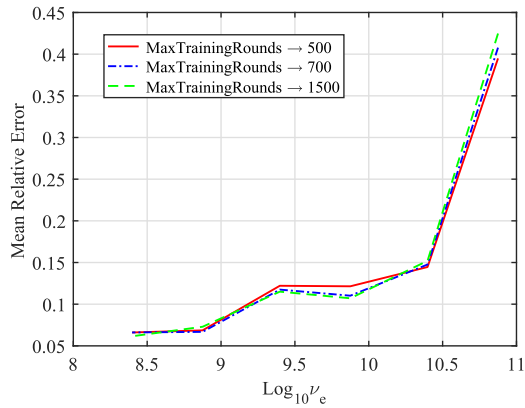

(c) MaxTrainingRounds

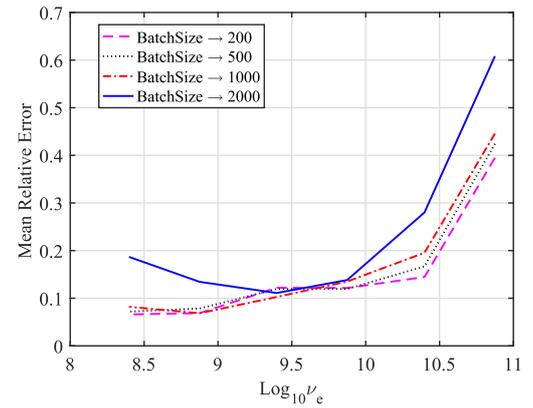

(d) BatchSize

**Figure 10.** (a)–(d) Represent mean relative error obtained by deep convolutional neural network for estimating frequency in the region $10^8$–$10^{11}$ Hz with different DropoutLayer, L2Regularization, MaxTrainingRounds, BatchSize respectively. Here, 12 000 training samples, 4000 validation samples and 4000 testing samples are used and they are disjoint.

## 4. Parameters estimation with deep learning

In this paper, we only focus on extra HFGWs generated by inflaton oscillations around the minimum of a cuspy potential after inflation in [28]. The extra HFGWs are distributed in amplitude region $10^{-36}$–$10^{-30}$ and frequency band $10^8$–$10^{11}$ Hz. In this section, we would estimate three parameters: dimensionless amplitude $h(t)$ and frequency $\nu_g = \omega_g/2\pi$ of extra HFGWs, and waist radius $W_0$ of GB. Considering mean relative error as the loss function, we make the parameters estimation using the similar CNN shown in figure 2.

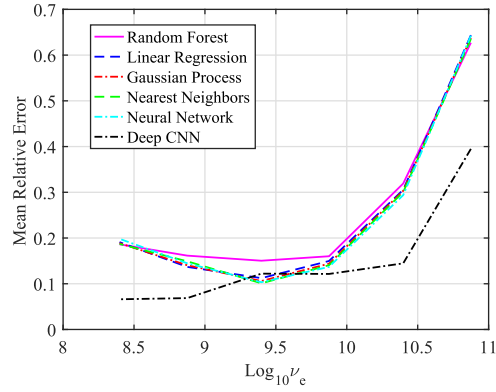

**Figure 11.** This is the mean relative error obtained by various machine learning algorithms for estimating frequency in region $10^8$–$10^{11}$ Hz. Here, 12 000 training samples, 4000 validation samples and 4000 testing samples are used and they are disjoint.

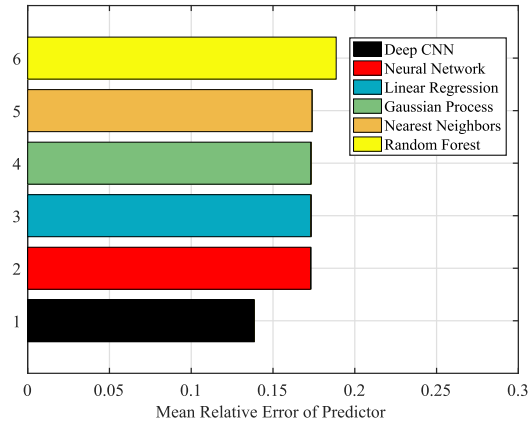

**Figure 12.** This is the mean relative error obtained by various machine learning algorithms for estimating waist radius in region 0.05–0.1 m. Here, 12 000 training samples, 4000 validation samples and 4000 testing samples are used and they are disjoint.

*For the estimation of GW amplitude:* the amplitude is an important characteristic property of GW, which affects on the possibility of recognition to a great extent. By adopting deep learning, we could find an optimal detection range of extra HFGWs. As the same process of tuning hyperparameters in classification, we get the corresponding results in figure 8. It can be found that the mean relative error is decreasing when the amplitude is increasing on the whole (see figure 8(a)). In the optimal case (i.e. DropoutLayer is set to be default) we further investigate how the L2Regularization, MaxTrainingRounds and Batchsize affect on it (see figures 8(b)–(d)). In summary, the predictor with DropoutLayer and L2Regularization omitted, MaxTrainingRounds and BatchSize set to be 700 and 100 respectively could compress the mean relative error less than 0.018 in whole amplitude region $10^{-36}$–$10^{-30}$.

Compared with predictors using common machine learning methods, such as Gaussian Process, k-Nearest Neighbors, Linear Regression, Neural Network and Random Forest, the predictor adopting deep CNN algorithm is more expressive. For our data sets, the predictor produced by deep CNN provides the lowest mean relative error in entire amplitude range $10^{-36}$–$10^{-30}$ as shown in figure 9, which is less than 0.016.

*For the estimation of GW frequency:* extra HFGWs have a broad frequency band $\nu_g \sim 10^8 - 10^{14}$ Hz. Here we concern on the frequency band from $10^8$ to $10^{11}$ Hz. On the whole the mean relative error of deep CNN increases with frequency increasing in $10^8$–$10^{11}$ Hz as shown in figure 10. In (a) of figure 10, the errors of these predictors are influenced by DropoutLayer obviously, and the optimal value of DropoutLayer should be default. So with the default of DropoutLayer, we make a further discussion about other hyperparameters. Finally it is found that the L2Regularization and MaxTrainingRounds have little effect on the performance of CNN. From (b) to (d) in figure 10, the predictor might be more suitable for our data sets with L2Regularization, MaxTrainingRounds and BatchSize set to be default, 500 and 200 respectively without DropoutLayer. Through adjusting these hyperparameters, the mean relative errors of predictors vary from 0.05 to 0.45 in frequency band $10^8$–$10^{11}$ Hz. At the optimal resonance frequency of extra HFGWs detector $5 \times 10^9$ Hz [30], the mean relative error of optimal deep CNN is around 0.12.

In the same way, a comparison between deep CNN and the predictors using other machine learning methods is shown in figure 11. Our predictor is able to successfully measure the frequency for given noisy signals with a relatively lower error.

*For the estimation of waist radius of GB*: waist radius $W_0$ describes minimum spot radius of GB. We investigate the mean relative error of waist radius from 0.05 m to 0.1 m by using deep CNN and other machine learning algorithms. The exact result is shown in figure 12. Compared to other predictors, the mean relative error of deep CNN is the lowest, which is 0.138. Thus, there will be observable effects, as long as the waist radius is set in region (0.05 m, 0.1 m).

## 5. Conclusion and remarks

Through the interaction between static magnetic field and extra HFGWs generating transverse first-order perturbative EM field, the transverse PPF as a special EM effect from extra HFGWs produced by inflation oscillations around the minimum of a cuspy potential after inflation could be generated. The amplitude of extra HFGWs is in the region ($10^{-36}$, $10^{-30}$) and the frequency ranges from $10^8$ to $10^{11}$ Hz. Through the application of deep CNN for extracting GW signal and the corresponding parameters estimation, the efficiency of deep learning technology has been suggested. Deep CNN can be trained by a large number of training datasets and is efficient to do classification and prediction after training appropriately. In training process, we use DropoutLayer and L2Regularization to avoid overfitting. Our deep CNN could be successfully classify and predict for our data sets. In this paper, we also discuss how the ratio of positive and negative samples affects on classification accuracy of classifier. One can find that the accuracy could reach up to 97.63%, when the ratio exceeds to 0.03. Especially, when the ratio exceeds to 0.11, the accuracy can reach up to 100%. Moreover, the classification accuracy in whole amplitude region $10^{-36}$–$10^{-30}$ shall be close to 100% with training 12000 training sets. Through analysis of some hyperparameters, including DropoutLayer, L2Regularization, MaxTrainingRounds and BatchSize, the accuracy of classifier from deep CNN is higher than the other commonly used classifiers in whole amplitude range. Therefore, extra HFGWs would be possible to be extracted from raw noisy datasets with high confidence level by deep CNN. The GW with stronger amplitude resulting in relatively high SNR, which is much easier to be recognized by such scheme. The mean relative error decreases when the amplitude of extra HFGWs increases. Signals with amplitude ranging from $10^{-31}$ to $10^{-30}$ are easier to be recovered. Fortunately, the amplitude of HFGWs predicted by several classical cosmological models are in this region, such as the quintessential inflationary models, some string cosmology scenarios and nano piezoelectric crystal array. Through tuning some hyperparameters, the optimal architecture of deep CNN can be fixed. One can find that both the classifier and predictor from deep CNN have better performance than traditional machine learning methods. Therefore, the PPF generated from extra HFGWs could be distinguished from one-dimensional noisy data sets with high confidence level by deep learning. Our results indicate that the deep learning technique could help us to improve the operability of extra HFGWs classification and parameters estimation.

## Acknowledgments

This work has been supported by the National Natural Science Fund of China (Grant No. 11873001, 11775038 and 11847301), and by the Fundamental Research Funds for the Central Universities (Grant No. 106112017CDJXFLX0014 and 2019CDJDWL0005), and by the Nature Science Fund of Chongqing No. cstc2018jcyjAX0767.

## ORCID iDs

Jin Li  https://orcid.org/0000-0001-8538-3714